# THE USE OF ASTRONOMICAL DATA FOR SATELLITE TRACKING

**Rositsa Miteva[1], Nikola Antonov[2,3], Adrian Sonka[4]**

[1]*Institute of Astronomy and National Astronomical Observatory – Bulgarian Academy of Sciences*
[2]*Astronomical Observatory Meshtitsa, Institute for Advanced Physical Studies*
[3]*Faculty of Physics, Sofia University "St. Kliment Ohridski"*
[4]*Astronomical Institute of the Romanian Academy*
*e-mail: rmiteva@nao-rozhen.org*

***Abstract:*** *The study explored the usage of astronomical observations for the identification and tracking of artificial satellites. Spacecraft streaks on astronomical images are a growing issue for the astronomical community. The increasing number of satellites in the future will only worsen the situation for ground-based optical and radio observations of cosmic objects. In addition, the spacecraft passages often lead to the discarding of the obtained data. In this analysis, we propose an estimation of the usefulness of dedicated astronomical observation for spacecraft monitoring, identification, and deduction of orbital parameters. We use astronomical data from the Astronomical Observatory Mestitsa (Bulgaria) and the https://www.tycho-tracker.com/ software for the streak analysis. The results are compared with well-known satellite databases, such as https://celestrak.org/. The findings are discussed in the framework of space weather research.*

# ОЦЕНКА ЗА ПРИНОСА НА АСТРОНОМИЧЕСКИ ДАННИ
# ПРИ ПРОСЛЕДЯВАНЕ НА СПЪТНИЦИ

**Росица Митева[1], Никола Антонов[2,3], Адриан Сонка[4]**

[1]*Институт по астрономия с Национална астрономическа обсерватория –
Българска академия на науките*
[2]*Астрономическа обсерватория Мещица, Институт за съвременни физически изследвания*
[3]*Фичически факултет, Софийски университет "Св. Климент охридски"*
[4]*Астрономически университе на Румънската академия*
*e-mail: rmiteva@nao-rozhen.org*

***Ключови думи:*** *изкуствени спътници, астрономически наблюдения, орбитални параметри*

***Резюме:*** *Това изследване предлага оценка за приноса на астрономически наблюдения за идентификацията и проследяването на изкуствени спътници. Следите от спътници върху астрономически изображения са все по-чест проблем за астрономите. Увеличаващият се брой на спътници и в бъдеще само ще вложи настоящата ситуация за наземните оптични и радио наблюдения на космически обекти. В този анализ ние предлагаме оценка на ползата от специализирани астрономически наблюдения за проследяването на спътници, тяхното разпознаване и определяне на орбитални параметри. За целта използваме астрономически данни от Астрономическа обсерватория Мещица (България) и софтуера https://www.tycho-tracker.com/ за идинтифициране на следите от спътници. Резултатите са сравнени с данни от добре известни спътникови бази-данни като https://celestrak.org/. Обсъдено е и приложението им към тематиката на космическото време.*

**Introduction**

The accurate and uninterrupted satellite monitoring has widespread technological and practical applications in everyday life. The satellite industry continues to grow with an immense number of small-scale commercial satellites, mostly in low Earth orbit (LEO), e.g., the Starlink constellation, providing internet connection worldwide.

Apart from the natural orbital decay or/and collisions, satellites are subject to the changes in their plasma environment, both in the interplanetary (IP) medium and in the terrestrial magnetosphere or atmosphere, depending on the orbit. These changes are driven mostly by the variability of the Sun, e.g., increased photon emission, particle and plasma eruptions, transported via the solar wind through the IP space and the environment around the planet. These solar-driven drivers can lead to detectable effects in the ground-based or space-borne technological devices and pose a threat to human health/life. This is the concept of space weather (Temmer, 2021).

Isolated cases of spacecraft malfunctions, up to and including spacecraft loss, have been reported in the literature, e.g., a short summary can be found in Miteva et al. (2023). Among the most probable reasons are increased flux of solar radiation, energetic particles, and geomagnetic storms (GSs), the latter being the decrease in the ground-based magnetic field as a response to a magnetized plasma bubble hitting the terrestrial magnetosphere. One way to denote the strength of the GS is by the so-called disturbance storm index (Dst), measured in negative nanoTesla (https://isgi.unistra.fr/indices_dst.php).

A recent example of the GS effects on spacecraft orbital decay was demonstrated by Wu et al. (2025). The authors calculated the orbital decay by up to several hundred meters during the course of several GSs. Thus, one could explore the GS effects on spacecraft orbital parameters by preselecting periods of an ongoing GS.

Instead of using standard sources for satellite positioning, a different approach is proposed in this study, in terms of searching for an alternative source of satellite observations. During long exposure observations of the sky, satellites leave a track while passing in the field-of-view (FOV) of telescopes. Satellite streaks/trails on astronomical images in the optical (Kruk et al. 2023) and also in the radio (Di Vruno et al. 2023) domains are nowadays recognized as a serious problem for astronomy. Most commonly, such observations are regarded as compromised and are often discarded.

Alternatively, one might focus on the streaks instead, as the added value of dedicated astronomical observations has not been considered or evaluated in full, especially under the framework of satellite monitoring campaigns. This is why, in this study, the spacecraft streaks in the FOV of astronomical images are used (1) for satellite identification and (2) to deduce the payload/debris flight parameters. This study evaluates what amount of complementary information for spacecraft monitoring could be provided by a single, ground-based observing station, as in the case of Astronomical Observatory (AO) Meshtitsa in Bulgaria (Antonov, 2025).

**Event selection**

In order to evaluate the usefulness of astronomy-dedicated observations, we scanned the entire database of an AO Meshtitsa in Bulgaria, https://astro.iaps.institute/. The AO is operational from late 2023. The observatory is entirely remote and aims at observing selected FOVs on the sky, when weather permitting. For this study, we browsed through all collected observations over the entire year 2024. A dedicated software was written to automatically preselect all images where satellite streaks are present. The routine returned the following results for the number of astronomical images with satellite passing per month during 2024, namely:

- Jan: 5
- Feb: 5
- Mar: 13
- Apr: 96
- May: 10
- Jun: 91
- Jul: 170
- Aug: 97
- Sep: 58
- Oct: 28
- Nov: 9
- Dec: 11

leading to 593 individual images in total. The number of streak observations varies due to, e.g., weather conditions/seasons, pre-selected FOV with targets of observations, and duration of observations.

In the next phase of the analyses, the geomagnetic disturbances in 2024 were collected, using the Kyoto-database:

- provisional Dst index, https://wdc.kugi.kyoto-u.ac.jp/dst_provisional/index.html.

Among all disturbances, we filter out the following 12 strong GSs with Dst < −100 nT, listed below per month during 2024:

- Mar: 3rd, 20 UT, −112 nT; 24th, 21 UT, −128 nT
- **Apr: 19th, 20 UT, −117 nT**
- May: 11th, 3 UT, −406 nT
- Jun: 28th, 13 UT, −107 UT
- **Aug: 4th, 18 UT, −100 UT; 12th, 17 UT, −188 nT**
- **Sep: 12th, 15 UT, −121 nT**; 17th, 6 UT, −120 nT
- Oct: 8th, 8 UT, −148 nT; 11th, 2 UT, −333 nT
- **Nov: 9th, 13 UT, −101 nT**

where the five dates denoted in bold font contain also observed satellite streaks in the considered here astronomical data, or 18 individual images in total. As the GS lasts for many hours, even if the peak of the storm is recorded during local daytime and the astronomical observations are done during local nighttime, we will keep the case for further investigation.

In summary, routine astronomical observations from AO-Mestitsa coincided with strong GSs in 5 out of 12 cases. Also, only 18 out of all 593 FITS files with automatically pre-identified satellite streaks were recorded from a single location during strong GSs over the year 2024.

**Methodology**

As stated above, the goal of this study is to explore the possible effects on the satellite trajectories due to space weather events, especially during the course of a GS. As a demonstration of the analyses, we selected the case observed during the strongest GS in the list, namely 2024-08-12. The monthly values of the Dst index (provisional) are shown in Fig. 1 with 1-hr resolution. The minimum Dst value of −188 nT is recorded at 17 UT. The GS shows a double dip profile, with the first minimum of −139 nT at 8 UT.

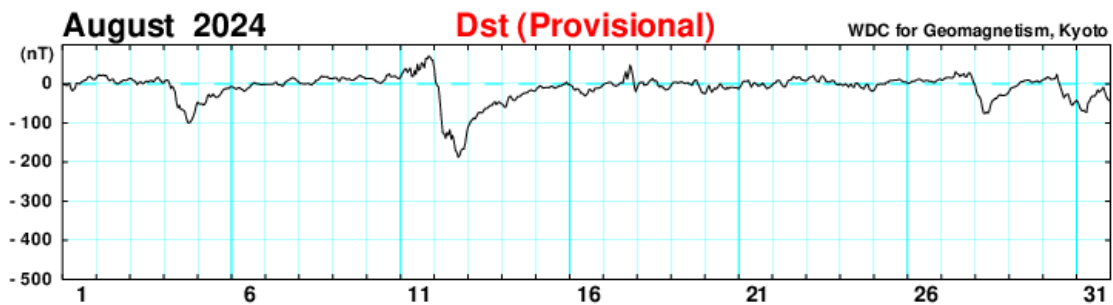

Fig. 1. Monthly trend of the Dst index with 1-hr resolution adopted from https://wdc.kugi.kyoto-u.ac.jp/dst_provisional/202408/index.html

Using a dedicated software, https://www.tycho-tracker.com/, we proceeded to analyze the data, focusing on the examples with a visible satellite track. The software loads the FITS file, and the FOV can be easily visualized, as shown in Fig. 2, focused on the Triangulum (Tri) constellation in this particular case. The initial and final points of the track are selected by hand and then supplied to the software, which returns a set of orbital parameters, including the satellite identification.

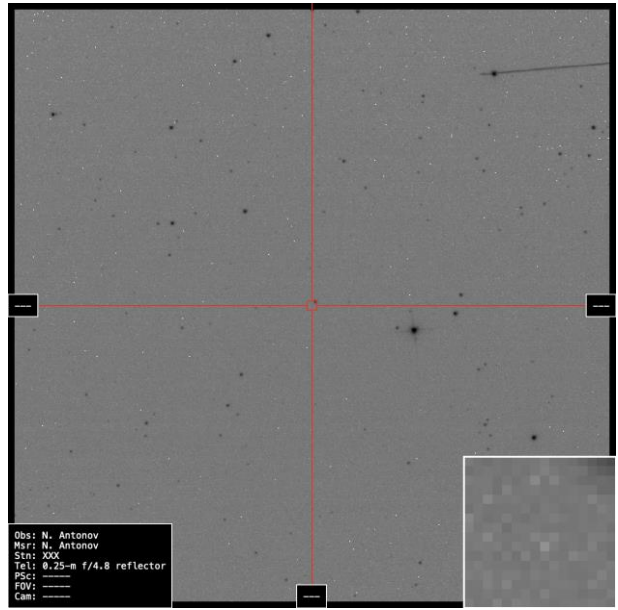

Fig. 2. An example of an astronomical image from AO-Mestitsa with a satellite streak in the top right quadrant during the GS of 2024-08-12

The next phase of the analysis is to compare the Tycho Tracker findings with the available information from independent satellite databases, among which we use the following:

- https://www.n2yo.com/
- https://celestrak.org/satcat

**Results**

For the example case of the track observed on 2024-08-12 (Fig. 2), the Tycho Tracker software provided the following results:

- object name: HJS-4A (HULIANWANG JISHU SHIYAN)
- object number: 58691U
- object speed (predicted speed of the matched object): 50610.82
- object distance (cross-track distance of the matched object, in arcminutes): 14.506

The type of this object is a payload. The object name and number are confirmed by the following databases. Firstly, the database https://www.n2yo.com/satellite/?s=58691#results confirms not only the satellite identity, but also provides the following additional information:

- NORAD ID: 58691
- International Code: 2023-212A
- Perigee: 1,110.5 km
- Apogee: 1,111.7 km
- Inclination: 50.0°
- Period: 107.3 minutes
- Semi-major axis: 7482 km
- Launch date: December 30, 2023
- Source: People's Republic of China (PRC)
- Launch site: Jiuquan Satellite Launch Center, China (JSC)

The second database is CelesTrak's satellite catalogue (SATCAT) search results: https://celestrak.org/satcat/table-satcat.php?CATNR=58691&PAYLOAD=1&ACTIVE=1&ORBIT=1&MAX=500. Namely, the trends of the orbital parameters during the entire month of August 2024 provided from https://celestrak.org/ are shown in Fig. 3. Among the plotted orbital parameters, only a slight increase of the eccentricity is observed at the time of the GS (2024-08-12), however such increases/decreases are no isolated feature and occur during quiet periods as well (e.g., during the period between 6th and 10th August, see Fig. 3).

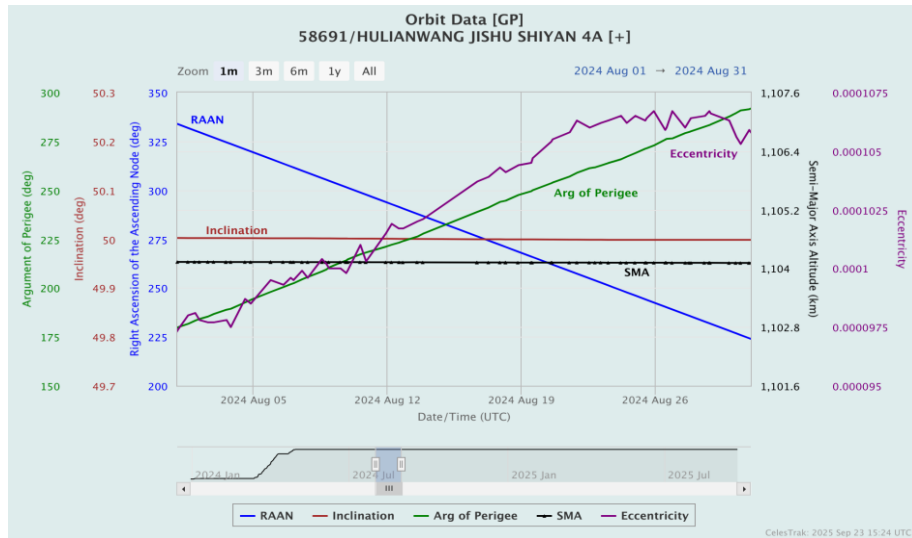

Fig. 3. Orbital parameters of HJS-4A during August 2024 adopted from
https://celestrak.org/NORAD/elements/graph-orbit-data.php?CATNR=58691

In a similar way, we analysed the remaining events from the event list (18 in total). The results are listed below:

- In 7/18 cases, the satellite identification from multiple sources did not provide consistent results, leading to about 61% success rate of the procedure. All but one of the failed identifications are debris.
- In 8/18 cases, the software reported decay or debris, whereas the remaining are payloads (for civilian and military use).
- Detected changes in satellite orbital parameters during the course of the GSs: among the pre-selected parameters from https://celestrak.org/ (argument of perigee (deg), inclination (deg), right-ascension of the ascending mode (deg), semi-major axis altitude (km), and eccentricity) it is the eccentricity that exhibits changes, however a causality cannot be confirmed at this stage.

In order to test the sensitivity of the reported orbital parameters, we compared the timings of the sudden orbital drop of FENGYUN 3G satellite by Wu et al. (2025) during the GS on 2024-05-11 (with Dst = −406 nT at 3 UT) with the parameter trends from the above database. In addition to the variations of the eccentricity, it is the semi-major axis altitude (in km) that showed a gradual, but notable change from its overall trend (namely a drop-like decline) starting on the day of the GS. Compared with the results for the strongest GS from our list, it is plausible that changes in the orbital parameters from https://celestrak.org/ can be noticed during extreme storms only.

In the future, we will employ an alternative approach, namely the observed location of these satellites and compare them to the Two Line Element (TLE) positions from space-track.org in order to quantify changes in the orbital parameters due to (weaker) GSs. The quantitative determination of the variations of the satellite orbital parameters as a function of the GS strength, duration, profile, etc. is the ultimate goal of such studies and will be of practical use for the pre-determination of satellite stability conditions during space weather events.

**Conclusions**

Based on the data analyses over one year (2024), we confirm that, overall, there is limited usage of astronomy-dedicated data for spacecraft monitoring due to multiple factors: limited FOV and UT data coverage of a single ground-based astronomical observatory, weather conditions, and/or a very limited number of strong GSs during the ongoing solar cycle 25. In addition, there is a limited application of automatic routines to the data, apart from streak recognition. Also, there is a need for event post-analysis, which is often manual (satellite identification, object orbital parameters, and TLEs).

Nevertheless, astronomy-dedicated data have the potential to be useful (1) as a complementary and independent data source and (2) for historical event analyses, provided that data from a network of such ground-based stations is available. Single-point observations (by a given observatory) can cover a very limited fraction of all-sky satellite observations. A network of stations is needed with coordinated efforts triggered especially during a forecasted GS.